\documentclass[aps,prd,nofootinbib,superscriptaddress]{revtex4-2}
\usepackage{amsmath,amssymb,bm}
\usepackage{graphicx}
\usepackage[colorlinks=true,allcolors=blue]{hyperref}
\usepackage{booktabs}
\usepackage{siunitx}
\usepackage{placeins}

\newcommand{\dd}{\mathrm{d}}

\newcommand{\mpl}{M_{\rm Pl}}
\newcommand{\lcdm}{\ensuremath{\Lambda\mathrm{CDM}}}
\newcommand{\rhoc}{\rho_{\rm c}}
\newcommand{\rhophi}{\rho_{\phi}}
\newcommand{\pphi}{p_{\phi}}

\newcommand{\Omphi}{\Omega_{\phi}}
\newcommand{\rd}{r_{\rm d}}
\newcommand{\fsig}{f\sigma_8}
\newcommand{\cH}{\mathcal{H}}

\newcommand{\weff}{w_{\rm eff}}

\usepackage{mathtools}
\usepackage{xspace}
\usepackage{enumitem}

\newcommand{\ltit}{\textsc{LTIT}\xspace}
\newcommand{\class}{\textsc{CLASS}\xspace}
\newcommand{\camb}{\textsc{CAMB}\xspace}

\setlist[enumerate]{leftmargin=*,itemsep=2pt,topsep=3pt}
\allowdisplaybreaks[2]

\begin{document}

\title{Late-Transition Interacting Thawer Dark Energy: Model Definition and Benchmark Consistency Tests}

\author{Slava G. Turyshev}
\author{Diogo H. F. de Souza}
\affiliation{
Jet Propulsion Laboratory, California Institute of Technology,\\
4800 Oak Grove Drive, Pasadena, CA 91109-0899, USA
}%

\date{\today}

\begin{abstract}
We formulate the late-transition interacting thawer (LTIT) as a late-activating, variable-coupling realization of coupled quintessence in which a canonical scalar field couples conformally only to cold dark matter. The construction is a non-universal dark-sector interaction: baryons and radiation remain minimally coupled, while CDM follows a conformally related metric. This structure ties three observational sectors to one microscopic mechanism--the pre-recombination calibration scale, the low-redshift expansion history, and the coupled growth response. We derive the exact background equations, the exact CDM scaling identity $\rhoc(a)=\rho_{c0}a^{-3}C[\phi(a)]/C(\phi_0)$, and the coupled CDM--scalar perturbation equations in synchronous gauge in a form suitable for Einstein--Boltzmann solvers. In the representative benchmarks, $\Omega_\phi(z_*)\sim10^{-9}$, $|\Delta r_d/r_d|\le3.82\times10^{-3}$, and $\max_{0<z<3}|E/E_{\Lambda\mathrm{CDM}}-1|\le0.380\%$, while the quasi-static growth indicator ranges from $0.725\%$ to $5.94\%$. LTIT therefore demonstrates how late activation can preserve the early calibration sector while producing a perturbation-sensitive signature, providing a concrete target for Boltzmann-level tests of interacting dark energy.
\end{abstract}

\maketitle

\section{Introduction}
\label{sec:intro}

DESI-era dark-energy analyses have sharpened a specific physical question: when a combined BAO--SN--CMB analysis shows a small apparent departure from $w=-1$, is the signal a genuine low-redshift dynamical effect, a shift of the early calibration scale $\rd$, or a residual systematic mode in the joint likelihood? DESI DR2 BAO measurements, CMB calibration from Planck, and recent supernova samples provide complementary pieces of this comparison, while the published post-DESI synthesis emphasizes that evolving-dark-energy interpretations remain dataset- and parametrization-dependent, with supernova calibration and selection effects playing an important role \cite{DESIDR2BAO,Planck2018VI,Brout2022PantheonPlus,Abbott2024DESY5SN,Turyshev2026DESIStatus}. This situation calls for microphysical models that are more constrained than a background-only $w(z)$ fit and that can be tested simultaneously in geometry and growth.

The late-transition interacting thawer (\ltit) addresses this problem as a controlled late-time branch of coupled quintessence. A canonical scalar field is conformally coupled only to cold dark matter, and the coupling becomes appreciable only after the field reaches a threshold in field space. The construction is designed to answer a focused question: can a low-redshift dark-sector interaction deform $H(z)$ while leaving the sound-horizon calibration essentially intact and without introducing phantom microphysics?

The restrictive feature is that the same interaction controlling the homogeneous CDM density also acts on the coupled CDM perturbations. LTIT is therefore more predictive than a background parametrization such as CPL: once the coupling turns on, it changes $H(z)$, the CDM continuity equation, and the CDM Euler equation. The central thesis of this paper is that early-time protection, low-redshift expansion response, and growth-sector closure must be assessed together. We formulate the model, derive the background and synchronous-gauge perturbation equations, and use representative benchmarks to demonstrate the hierarchy of tiny early-time effects, sub-percent background deformations, and a larger late-time growth diagnostic. A likelihood-level confrontation with data is the next inference step and requires the solver-level implementation specified below.

The paper is organized as follows. Section~\ref{sec:model} defines the LTIT model and its field-theory structure. Section~\ref{sec:background} derives the background dynamics and the effective phantom condition. Section~\ref{sec:sync} presents the synchronous-gauge perturbation equations and initial conditions. Section~\ref{sec:justification} summarizes the physical interpretation and design criteria. Section~\ref{sec:bench} presents benchmark consistency tests. Section~\ref{sec:implementation} summarizes the solver-level requirements implied by the LTIT system. Section~\ref{sec:conclusion} concludes.

\section{LTIT as a variable-coupling coupled-quintessence model}
\label{sec:model}

\subsection{Action and Einstein-frame interpretation}

We work in the Einstein frame with metric $g_{\mu\nu}$ and reduced Planck mass $\mpl$. Baryons and radiation are minimally coupled. CDM is conformally coupled to a canonical scalar field $\phi$:
\begin{align}
S &= \int \dd^4x\sqrt{-g}\left[\frac{\mpl^2}{2}R - \frac12 (\nabla\phi)^2 - V(\phi)\right]
+ S_b[g_{\mu\nu},\psi_b] + S_r[g_{\mu\nu},\psi_r] + S_c[\tilde g_{\mu\nu},\psi_c],
\qquad \tilde g_{\mu\nu}=C^2(\phi) g_{\mu\nu}.
\label{eq:action}
\end{align}
The dimensionless coupling function is defined by
\begin{equation}
\beta(\phi) \equiv \mpl \frac{\dd \ln C(\phi)}{\dd \phi}.
\label{eq:beta_def}
\end{equation}
For constant $\beta$, Eqs.~(\ref{eq:action})--(\ref{eq:beta_def}) reduce to standard coupled quintessence \cite{Amendola2000CQ,Amendola2004Perturbations}. Field-dependent and time-dependent dark-sector couplings have been studied as controlled extensions of this framework, including their impact on structure formation \cite{PettorinoBaccigalupi2008CQ,Baldi2011TimeDependentCouplings}. LTIT is the late-transition realization in which the coupling is strongly suppressed until the field reaches a threshold.

In this paper, LTIT denotes this specific late-transition, field-dependent coupling ansatz: an early-time-protected subclass of coupled quintessence constructed to isolate late dark-sector dynamics from the calibration sector.

The coupling in Eq.~(\ref{eq:action}) is non-universal. It is a phenomenological dark-sector interaction in which CDM follows the conformally related metric $\tilde g_{\mu\nu}$, whereas baryons and radiation are assumed to remain minimally coupled to $g_{\mu\nu}$. It should therefore not be interpreted as the Einstein-frame form of a scalar--tensor theory with a universal Jordan-frame matter coupling. This distinction is important because a universal Jordan-frame coupling would generically induce scalar couplings to the full matter sector after a conformal transformation, while LTIT deliberately couples only the dark matter sector.

\subsection{Late-transition ansatz and closed-form conformal factor}

The defining LTIT ansatz is a smooth step in field space,
\begin{equation}
\beta(\phi)=\frac{\beta_0}{2}\left[1+\tanh\!\left(\frac{\phi-\phi_t}{\Delta\phi}\right)\right],
\label{eq:beta_tanh}
\end{equation}
with derivative
\begin{equation}
\beta_{,\phi}(\phi)=\frac{\beta_0}{2\Delta\phi}\,\mathrm{sech}^2\!\left(\frac{\phi-\phi_t}{\Delta\phi}\right).
\label{eq:beta_prime}
\end{equation}
The functions $C(\phi)$ and $\beta(\phi)$ are not independent. Specifying $\beta(\phi)$ fixes $C(\phi)$ up to a multiplicative normalization, and specifying $C(\phi)$ fixes $\beta(\phi)$. We use the normalization $C(\phi_0)=1$, which can be absorbed into the present-day CDM density normalization. Since Eq.~(\ref{eq:beta_def}) can be integrated explicitly, the conformal factor is
\begin{equation}
\ln\!\frac{C(\phi)}{C(\phi_0)}=
\frac{\beta_0}{2\mpl}
\left[
(\phi-\phi_0)+\Delta\phi\ln\!\frac{\cosh\!\left((\phi-\phi_t)/\Delta\phi\right)}{\cosh\!\left((\phi_0-\phi_t)/\Delta\phi\right)}
\right].
\label{eq:Cphi}
\end{equation}
Eq.~(\ref{eq:Cphi}) makes explicit that the late-transition ansatz is a particular field-dependent coupled-quintessence interaction rather than an additional dark-sector species.

\subsection{Benchmark potential and parameter regime}

For the benchmark study we adopt the shallow exponential potential
\begin{equation}
V(\phi)=V_0\exp\!\left(-\lambda\frac{\phi}{\mpl}\right),
\label{eq:Vexp}
\end{equation}
with $\lambda\ll1$. The exponential potential supplies a minimal thawing background in which the role of the late coupling can be isolated from additional structure in $V(\phi)$. LTIT itself is defined by the late-transition coupling rather than by Eq.~(\ref{eq:Vexp}); any sufficiently shallow thawing potential that leaves the field effectively frozen until late times would preserve the same qualitative mechanism \cite{Copeland2006DynamicsDE}. The benchmark choice therefore provides a controlled realization of the late-transition mechanism.

\subsection{Why the transition must be late}

A non-negligible dark-sector coupling during the drag-epoch or recombination era generically perturbs the pre-recombination expansion rate, shifts the sound horizon $\rd$, modifies the CMB acoustic scale, and changes the matter-radiation era growth history. LTIT is built to avoid that. The phenomenological purpose of the tanh activation is thus not ad hoc freedom for its own sake; it enforces a clean temporal separation between the well-calibrated early universe and any allowed late-time dynamics. In this sense LTIT should be regarded as a deliberately constrained, DESI-era-friendly subclass of variable-coupling coupled quintessence, rather than as a generic interacting-DE model.

\section{Background dynamics, exact identities, and effective phantom behavior}
\label{sec:background}

\subsection{Covariant transfer and background equations}

The interaction implied by the conformal coupling can be written covariantly as
\begin{equation}
\nabla_\mu T^{\mu\nu}_{(c)} = Q^\nu,
\qquad
\nabla_\mu T^{\mu\nu}_{(\phi)} = -Q^\nu,
\label{eq:Qnu_cov}
\end{equation}
with
\begin{equation}
Q^\nu = \frac{\beta(\phi)}{\mpl}\,\rhoc\,\nabla^\nu\phi.
\label{eq:Qnu}
\end{equation}
For a spatially flat FLRW background we obtain
\begin{align}
\dot\rhoc + 3H\rhoc &= Q,
\label{eq:rhoc_bg}\\
\dot\rhophi + 3H(\rhophi+\pphi) &= -Q,
\label{eq:rhophi_bg}
\end{align}
with
\begin{equation}
Q = \frac{\beta(\phi)}{\mpl}\,\rhoc\,\dot\phi.
\label{eq:Qbg}
\end{equation}
The scalar equation of motion is
\begin{equation}
\ddot\phi + 3H\dot\phi + V_{,\phi} = -\frac{\beta(\phi)}{\mpl}\rhoc,
\label{eq:KG_cosmic}
\end{equation}
and the Friedmann equation reads
\begin{equation}
3\mpl^2 H^2 = \rho_r + \rho_b + \rhoc + \rhophi,
\qquad
\rhophi = \frac{\dot\phi^2}{2}+V(\phi).
\label{eq:Friedmann}
\end{equation}
The scalar pressure and microphysical equation of state are
\begin{equation}
\pphi = \frac{\dot\phi^2}{2}-V(\phi),
\qquad
w_\phi = \frac{\pphi}{\rhophi} \ge -1.
\label{eq:wphi}
\end{equation}

\subsection{Effective equation of state and the sign condition for phantom-like behavior}

Defining an effective dark-energy equation of state through
\begin{equation}
\dot\rhophi + 3H(1+\weff)\rhophi = 0,
\end{equation}
one finds
\begin{equation}
\weff = w_\phi + \frac{Q}{3H\rhophi}.
\label{eq:weff_def}
\end{equation}
Therefore a canonical scalar can mimic $\weff<-1$ whenever
\begin{equation}
Q < -3H\rhophi(1+w_\phi).
\label{eq:phantom_condition}
\end{equation}
Eq.~(\ref{eq:phantom_condition}) is the central physical point: LTIT never requires a phantom kinetic term. Any effective phantom phase is produced by energy transfer from CDM to the scalar sector. For the sign convention adopted here and for the benchmark branch with $\dot\phi>0$, this requires $\beta(\phi)<0$ after the transition activates. The benchmark models therefore use negative $\beta_0$, so that $Q<0$ at late times and $\weff$ can dip below $-1$ while the canonical scalar still satisfies $w_\phi\ge -1$.

\subsection{Exact CDM identity}

Using Eq.~(\ref{eq:beta_def}), the CDM continuity equation can be integrated exactly:
\begin{equation}
\rhoc(a)=\rho_{c0}\,a^{-3}\frac{C[\phi(a)]}{C(\phi_0)}.
\label{eq:rhoc_identity}
\end{equation}
Eq.~(\ref{eq:rhoc_identity}) is an exact identity and should be used as a stringent implementation-level validation test. In practice, a numerical solution that violates Eq.~(\ref{eq:rhoc_identity}) at more than the $10^{-6}$ level is not under adequate control.

\subsection{Dimensionless background system in $N=\ln a$}

For numerical work it is convenient to evolve the autonomous system in e-fold time $N\equiv\ln a$ with
\begin{equation}
\phi_N \equiv \frac{\dd\phi}{\dd N},
\qquad
\nu \equiv \phi_N,
\end{equation}
so that
\begin{align}
\phi_N &= \nu,
\label{eq:phiN}\\
\nu_N &= -\left(3+\frac{H_N}{H}\right)\nu - \frac{V_{,\phi}}{H^2} - \frac{\beta(\phi)}{\mpl}\frac{\rhoc}{H^2},
\label{eq:uN}\\
(\rhoc)_N &= -3\rhoc + \frac{\beta(\phi)}{\mpl}\rhoc\,\nu.
\label{eq:rhocN}
\end{align}
With $\rho_b\propto a^{-3}$ and $\rho_r\propto a^{-4}$, the Friedmann relation becomes
\begin{equation}
H^2 = \frac{\rho_b+\rho_r+\rhoc+V(\phi)}{3\mpl^2-\nu^2/2}.
\label{eq:H2N}
\end{equation}
Eqs.~(\ref{eq:phiN})--(\ref{eq:H2N}) define the exact background benchmark system used below.

\subsection{Early-time protection and its quantitative diagnostics}

LTIT is only useful if it does not significantly disturb the early-universe calibration sector. Two diagnostics are particularly important:
\begin{equation}
\frac{\Delta\rd}{\rd} \equiv \frac{\rd^{\rm LTIT}-\rd^{\lcdm}}{\rd^{\lcdm}},
\qquad
\rd = \int_{z_d}^{\infty}\frac{c_s(z)}{H(z)}\,\dd z,
\label{eq:rdratio}
\end{equation}
and the scalar fraction at recombination,
\begin{equation}
\Omphi(z_*) \equiv \frac{\rhophi(z_*)}{3\mpl^2 H^2(z_*)}.
\label{eq:Ophistar}
\end{equation}
Here $z_*$ denotes the recombination redshift, $z_*\sim 10^3$ for the benchmark cosmologies. In the benchmark models below, $\beta(z_*)\simeq 0$, $\Omphi(z_*)\sim10^{-9}$, and $|\Delta\rd/\rd|<4\times10^{-3}$, so the LTIT deformation is genuinely late-time.

\section{Gauge-fixed linear perturbations in synchronous gauge}
\label{sec:sync}

\subsection{Metric and conventions}

We adopt synchronous gauge because it matches standard Einstein--Boltzmann conventions and makes future solver implementations straightforward. We follow Ma--Bertschinger notation \cite{MaBertschinger1995} and write
\begin{equation}
\dd s^2 = a^2(\tau)\left[-\dd\tau^2 + (\delta_{ij}+h_{ij})\dd x^i\dd x^j\right],
\end{equation}
with scalar-mode decomposition
\begin{equation}
h_{ij}(\mathbf{k},\tau)=\hat k_i\hat k_j h(\mathbf{k},\tau)+6\left(\hat k_i\hat k_j-\frac13\delta_{ij}\right)\eta(\mathbf{k},\tau).
\end{equation}
Conformal-time derivatives are denoted by primes, and $\cH\equiv a'/a$.

\subsection{Background equations in conformal time}

The background equations become
\begin{align}
\rhoc' + 3\cH\rhoc &= +\frac{\beta(\phi)}{\mpl}\rhoc\phi',
\label{eq:rhoc_conf}\\
\phi'' + 2\cH\phi' + a^2V_{,\phi} &= -a^2\frac{\beta(\phi)}{\mpl}\rhoc.
\label{eq:KG_conf}
\end{align}
These define the sign convention used below.

\subsection{Synchronous-gauge scalar perturbations}

A consistent synchronous-gauge implementation of the coupled CDM and scalar sector is
\begin{align}
\delta_c' &= -\theta_c - \frac{h'}{2} + \frac{\beta(\phi)}{\mpl}\delta\phi' + \frac{\beta_{,\phi}(\phi)}{\mpl}\phi'\delta\phi,
\label{eq:deltac_sync}\\
\theta_c' &= -\cH\theta_c + \frac{\beta(\phi)}{\mpl}\left(k^2\delta\phi - \phi'\theta_c\right),
\label{eq:thetac_sync}\\
\delta\phi'' &+ 2\cH\delta\phi' + \left[k^2 + a^2V_{,\phi\phi} + a^2\frac{\beta_{,\phi}(\phi)}{\mpl}\rhoc\right]\delta\phi + \frac{h'}{2}\phi' 
\nonumber\\
&= -a^2\frac{\beta(\phi)}{\mpl}\rhoc\,\delta_c.
\label{eq:deltaphi_sync}
\end{align}
The scalar contributes to the total density, pressure, and momentum perturbations through
\begin{align}
\delta\rho_\phi &= \frac{\phi'\delta\phi'}{a^2} + V_{,\phi}\delta\phi,
\label{eq:drhophi}\\
\delta p_\phi &= \frac{\phi'\delta\phi'}{a^2} - V_{,\phi}\delta\phi,
\label{eq:dpphi}\\
(\rhophi+\pphi)\theta_\phi &= \frac{k^2}{a^2}\phi'\delta\phi,
\label{eq:thetaphi}
\end{align}
with vanishing scalar anisotropic stress. The Einstein equations remain
\begin{align}
k^2\eta - \frac12\cH h' &= 4\pi G a^2\delta\rho_{\rm tot},
\label{eq:ein1}\\
k^2\eta' &= 4\pi G a^2\sum_i(\rho_i+p_i)\theta_i,
\label{eq:ein2}\\
h'' + 2\cH h' - 2k^2\eta &= -8\pi G a^2\delta p_{\rm tot},
\label{eq:ein3}\\
(h+6\eta)'' + 2\cH(h+6\eta)' - 2k^2\eta &= -24\pi G a^2\sum_i(\rho_i+p_i)\sigma_i.
\label{eq:ein4}
\end{align}
Eqs.~(\ref{eq:deltac_sync})--(\ref{eq:ein4}) define the gauge-fixed linear system proposed for LTIT solver implementation.

\subsection{Initial conditions}

Because LTIT enforces $\beta\to 0$ at early times, adiabatic initial conditions reduce to the standard uncoupled thawer limit. Deep in radiation domination one may initialize the scalar sector as
\begin{equation}
\phi(\tau_i)=\phi_i,
\qquad
\phi'(\tau_i)\simeq 0,
\qquad
\delta\phi(\tau_i)\simeq 0,
\qquad
\delta\phi'(\tau_i)\simeq 0,
\label{eq:ICphi}
\end{equation}
while the fluid perturbations satisfy the usual synchronous-gauge adiabatic relations,
\begin{equation}
\delta_c=\delta_b=\frac34\delta_\gamma=\frac34\delta_\nu=-\frac12 h + \mathcal{O}(k^2\tau^2),
\qquad
\theta_c,\theta_b = \mathcal{O}(\beta,k^2\tau^3).
\label{eq:ICadiabatic}
\end{equation}
In practice one chooses $z_i$ so that $\beta(z_i)$ is numerically negligible and reuses the standard adiabatic initial-condition machinery of the Boltzmann solver for the fluid sector. In the benchmark integrations we fix the field origin by convention and take $\phi_i=0$. For the exponential benchmark potential, a constant shift of $\phi$ can be absorbed into $V_0$ and $\phi_t$, so only the displacement relative to the transition location is physically relevant.

These initial conditions are valid only while the interaction is negligible. Once the coupling activates, CDM is no longer geodesic and the synchronous-gauge CDM-comoving condition $\theta_c=0$ ceases to be admissible. Any implementation that freezes $\theta_c$ after activation is solving a different system.

\section{Physical interpretation and closure logic}
\label{sec:justification}

The preceding sections define LTIT as a restricted late-time deformation of coupled quintessence rather than as a free interacting-dark-energy parametrization. A constant or early-acting coupling would feed directly into the drag epoch and recombination dynamics, shifting $\rd$, altering the CMB acoustic scale, and modifying the matter-era growth history. The field-dependent activation in Eq.~(\ref{eq:beta_tanh}) is therefore the mechanism that separates the well-tested early universe from the late dark sector the model is intended to probe.

This structure also clarifies why LTIT is more informative than a background parametrization such as CPL. A background fit can reproduce a smooth low-redshift trend in $H(z)$, but it does not specify how energy-momentum is transferred, whether the underlying scalar sector is microphysically acceptable, or what happens to the CDM perturbations. In LTIT those questions are explicit. The scalar remains canonical, so $w_\phi\ge -1$, while an apparent $\weff<-1$ phase arises only through the transfer term in Eq.~(\ref{eq:weff_def}). The same interaction that alters the background also modifies the CDM continuity and Euler equations, Eqs.~(\ref{eq:deltac_sync})--(\ref{eq:thetac_sync}), so the model cannot improve distance data without paying a perturbation-level price.

The relevant standard of success is therefore not whether LTIT can imitate a CPL-like background trend, but whether it can do so while preserving early-time calibration and surviving the stronger consistency test from growth and lensing. That closure logic is the central physics argument of the paper.

\section{Benchmark consistency tests}
\label{sec:bench}

The benchmark section tests the LTIT design logic in the same order as the model is constructed. Table~\ref{tab:bench_phys} and Fig.~\ref{fig:earlyprotect} quantify early-time protection, Fig.~\ref{fig:betaweff} displays the late activation and transfer-driven effective equation of state, and Table~\ref{tab:bench_obs} with Fig.~\ref{fig:observable_space} compares the background and growth responses. The numerical conventions and background system used to generate these diagnostics are collected in Appendix~\ref{app:numerics}, so each displayed quantity is tied to a specified set of evolution equations. The background curves and early-time diagnostics are obtained by integrating the exact background equations. To connect the exact background to the retained original growth diagnostic, Fig.~\ref{fig:observable_space}(c) uses the single-fluid quasi-static indicator. In this benchmark equation we use the same e-fold notation as in Appendix~\ref{app:numerics}: $N\equiv\ln a$, $D_N\equiv\dd D/\dd N$, and $D_{NN}\equiv\dd^2D/\dd N^2$. The diagnostic is
\begin{equation}
D_{NN}+
\left(2+\frac{H_N}{H}\right)D_N
-
\frac32\Omega_m(a)\,\mu(a)D\simeq0,
\qquad
\mu(a)\simeq 1+2\beta^2(\phi)\frac{\Omega_c(a)}{\Omega_m(a)} .
\label{eq:growth_proxy}
\end{equation}
The $N$-subscript convention in Eq.~(\ref{eq:growth_proxy}) is intentionally distinct from the conformal-time primes used for the synchronous-gauge perturbation system in Sec.~\ref{sec:sync}. This equation is a diagnostic approximation only. Because the scalar couples directly to CDM but not to baryons, the exact perturbation problem cannot in general be reduced to a single scale-independent total-matter growth equation. The growth curve should therefore be interpreted as a scale-setting indicator for the expected sign and approximate magnitude of the response, not as a prediction for observed $f\sigma_8$, weak-lensing, CMB-lensing, or cluster observables.

The benchmark calculations establish internal consistency and quantify the intended hierarchy between early-time protection, sub-percent background deformation, and a potentially larger late-time growth response. They provide quantitative regression targets for a future Einstein--Boltzmann implementation. Full statistical inference requires implementing Eqs.~(\ref{eq:deltac_sync})--(\ref{eq:ein4}) and fitting background and perturbation-sensitive data jointly.

\subsection{Benchmark models and quantitative summary}

We define two representative benchmark points, both using the same transition location and potential slope but different asymptotic coupling amplitude. Benchmark A is intended to be conservative; Benchmark B illustrates a more aggressive but still early-time-safe case. Table~\ref{tab:bench_phys} specifies the microscopic parameters and the early-time diagnostics that test whether the construction leaves the calibration sector intact. Table~\ref{tab:bench_obs} then translates the same benchmarks into maximum background and quasi-static growth departures from matched flat \lcdm.

\begin{table*}[t]
\caption{\label{tab:bench_phys}
Microscopic benchmark parameters and early-time diagnostics. Here $z_{1/2}$ denotes the redshift at which the coupling reaches half its asymptotic amplitude, $\beta(z_{1/2})=\beta_0/2$. The key point is that the transition remains genuinely late: the scalar fraction at recombination is tiny and the induced shift in the sound horizon remains below the percent level.}
\centering
\setlength{\tabcolsep}{5pt}
\renewcommand{\arraystretch}{1.10}
\begin{tabular}{lccccccccc}
\toprule
Bench. & $\beta_0$ & $\phi_t$ & $\Delta\phi$ & $\lambda$ & $z_{1/2}$ & $\weff(0)$ & $\min\weff$ & $\Omphi(z_*)$ & $\Delta\rd/\rd$ \\
\midrule
A & $-0.20$ & $0.045$ & $0.002$ & $0.20$ & $0.122$ & $-0.9953$ & $-1.0000$ & $1.28\times10^{-9}$ & $-1.11\times10^{-3}$ \\
B & $-0.60$ & $0.045$ & $0.002$ & $0.20$ & $0.122$ & $-0.9997$ & $-1.0041$ & $1.27\times10^{-9}$ & $-3.82\times10^{-3}$ \\
\bottomrule
\end{tabular}
\end{table*}

\begin{table}[t]
\caption{\label{tab:bench_obs}
Maximum departures from matched flat \lcdm\ over the ranges indicated. The growth column is the quasi-static indicator in Eq.~(\ref{eq:growth_proxy}), not a full perturbation prediction.}
\centering
\renewcommand{\arraystretch}{1.10}
\begin{tabular}{lccc}
\toprule
Bench. & $\max_{0<z<3}|E/E_{\lcdm}-1|$ & $\max_{0<z<3}|(\fsig)/(\fsig)_{\lcdm}-1|$ & $\max_{z>10}|E/E_{\lcdm}-1|$ \\
\midrule
A & $0.158\%$ & $0.725\%$ & $0.111\%$ \\
B & $0.380\%$ & $5.94\%$  & $0.384\%$ \\
\bottomrule
\end{tabular}
\end{table}

The two benchmarks bracket a modest and a stronger late interaction. Benchmark A stays very close to \lcdm\ at all redshifts, with $\max_{0<z<3}|E/E_{\lcdm}-1|=0.158\%$ and $\max_{0<z<3}|(\fsig)/(\fsig)_{\lcdm}-1|=0.725\%$. Benchmark B amplifies the same mechanism while remaining early-time safe, reaching $\min\weff=-1.0041$, $|\Delta\rd/\rd|=3.82\times10^{-3}$, $\max_{0<z<3}|E/E_{\lcdm}-1|=0.380\%$, and a quasi-static growth shift of $5.94\%$. The numerical pattern is the central result: late activation protects the calibration sector while allowing a growth diagnostic that is an order of magnitude more responsive than the background expansion.

\subsection{Coupling history and effective equation of state}

\begin{figure}[!htbp]
\centering
\includegraphics[width=0.44\textwidth]{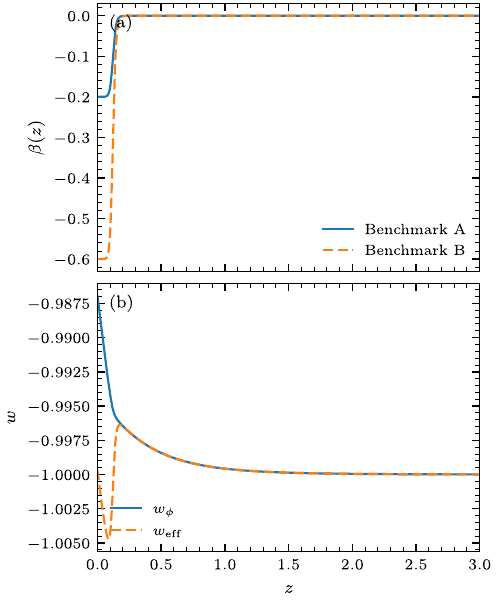}
\caption{\label{fig:betaweff}
Late activation and transfer-driven effective phantom behavior. Panel (a) shows $\beta(z)$ for the two benchmark models. Panel (b) shows benchmark B, for which the microphysical scalar equation of state $w_\phi$ remains above $-1$ while the effective quantity $\weff$ dips slightly below $-1$ after the interaction turns on. The excursion is therefore generated by energy transfer rather than by phantom microphysics.}
\end{figure}

Figure~\ref{fig:betaweff} isolates the mechanism that makes LTIT distinctive. The upper panel locates the activation in redshift, while the lower panel makes the sign logic derived in Sec.~\ref{sec:background} explicit: because $w_\phi\ge -1$ for a canonical scalar, the dip of $\weff$ below $-1$ must be transfer-driven. In benchmark B the excursion is mild, $\min\weff=-1.0041$, but it is sufficient to demonstrate that LTIT can generate effective phantom behaviour without a ghost degree of freedom.

\subsection{Early-time protection}

\begin{figure}[!htbp]
\centering
\includegraphics[width=0.44\textwidth]{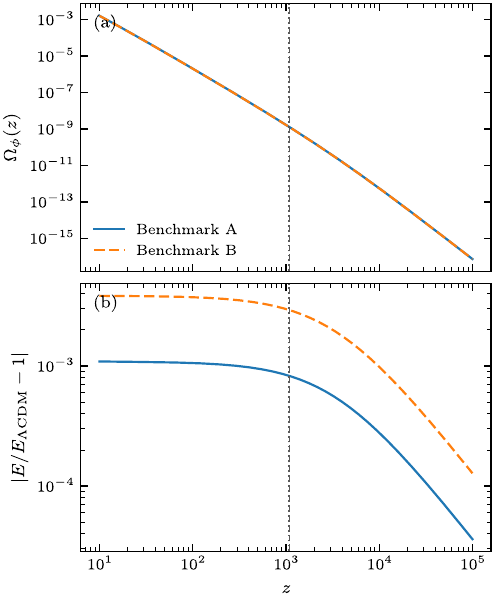}
\caption{\label{fig:earlyprotect}
Early-time protection in LTIT. Panel (a) shows the scalar fraction $\Omphi(z)$, which remains negligible at recombination and earlier epochs. Panel (b) shows the absolute fractional deviation of the high-redshift expansion rate from matched flat \lcdm. The benchmark deformation is therefore genuinely late-time rather than a hidden modification of the calibration sector.}
\end{figure}

Figure~\ref{fig:earlyprotect} addresses the main consistency requirement of the construction. For the benchmarks in Table~\ref{tab:bench_phys}, the scalar fraction remains at the level of $\Omphi(z_*)\sim10^{-9}$ and the induced sound-horizon shift stays below $4\times10^{-3}$. In other words, the late interaction does not secretly move the calibration sector; the model becomes dynamically relevant only after recombination and well after the drag epoch.

\subsection{Background and quasi-static growth diagnostics}

\begin{figure}[!htbp]
\centering
\includegraphics[width=0.72\textwidth]{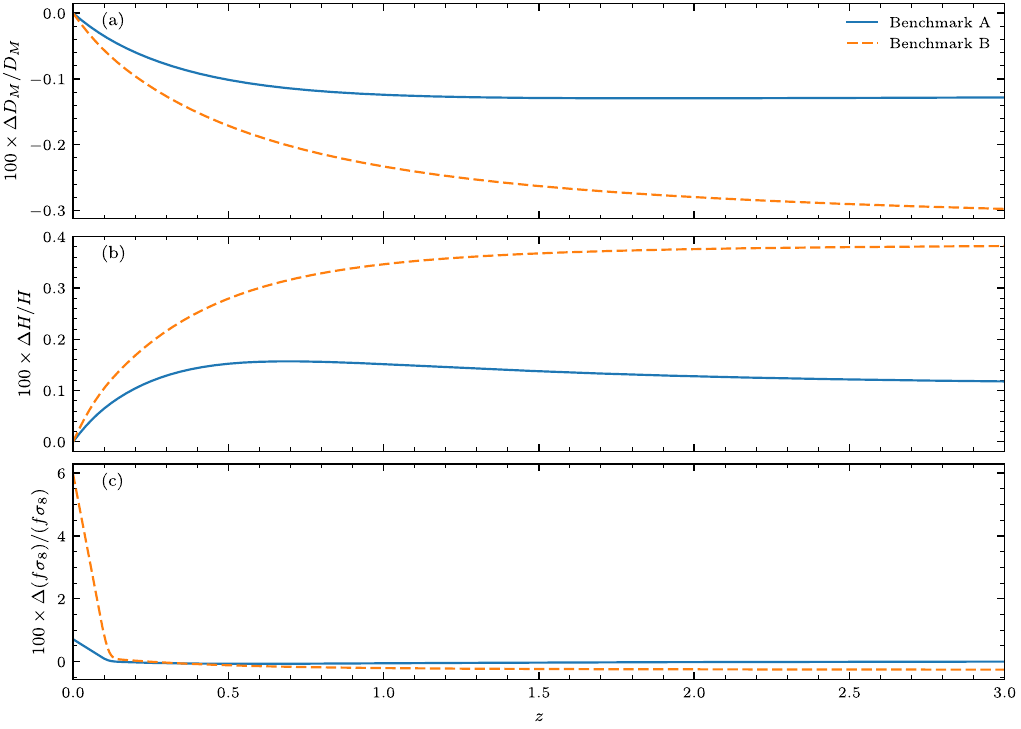}
\caption{\label{fig:observable_space}
Observable-space response relative to matched flat \lcdm. Panels (a)--(c) show $100\times\Delta D_M/D_M$, $100\times\Delta H/H$, and $100\times\Delta(f\sigma_8)/(f\sigma_8)$, respectively. Even when the background response stays below the percent level, the growth response can become appreciably larger after the interaction turns on. Background curves are exact; growth curves use the quasi-static approximation Eq.~(\ref{eq:growth_proxy}), with derivatives taken with respect to $N=\ln a$.}
\end{figure}

Figure~\ref{fig:observable_space} shows the observable-space benchmark diagnostics. The late-time background deformation remains deliberately small, while the quasi-static growth indicator is appreciably larger. In Benchmark B, $\max_{0<z<3}|E/E_{\lcdm}-1|<0.4\%$ while the quasi-static growth shift reaches $5.94\%$. The diagnostic therefore identifies the perturbation sector as the decisive arena for testing LTIT: a model nearly invisible in distance observables can still produce a measurable growth-sector response.

\section{Solver-level requirements}
\label{sec:implementation}

The LTIT equations are written in a form suitable for implementation in standard Einstein--Boltzmann solvers such as \class and \camb \cite{Lesgourgues2011CLASS,CLASSdocs,Lewis2000CAMB}. A correct LTIT implementation must evolve the scalar perturbation, the coupled CDM density and velocity perturbations, the minimally coupled baryon and radiation sectors, and the metric perturbations consistently. It must not restore the CDM-comoving synchronous condition once the interaction activates.

Before any numerical scan is trusted, the solver should reproduce the decoupling limit $\beta_0\to0$, satisfy the exact CDM identity Eq.~(\ref{eq:rhoc_identity}), recover the benchmark background diagnostics in Tables~\ref{tab:bench_phys} and \ref{tab:bench_obs}, and converge under changes of initial redshift, time step, and gauge convention. Cross-checks against an independent implementation are also important because interacting-dark-energy perturbation systems can be sensitive to sign conventions and gauge choices.

After these tests one can compute CMB spectra, BAO distances, RSD quantities, weak-lensing observables, CMB-lensing spectra, and cluster-count predictions with a controlled perturbation implementation. The benchmark figures in this paper provide regression tests and scale-setting diagnostics for that calculation, linking the model-building results directly to the observables required for a future data analysis.

\section{Conclusions}
\label{sec:conclusion}

We have defined LTIT as a late-transition, field-dependent, non-universal dark-sector realization of coupled quintessence. The model couples a canonical scalar conformally to CDM only, leaves baryons and radiation minimally coupled, and activates the interaction only after the field reaches a low-redshift threshold. The conformal factor $C(\phi)$ and the logarithmic coupling $\beta(\phi)$ are related by Eq.~(\ref{eq:beta_def}); in the benchmark study we specify $\beta(\phi)$ and use the corresponding closed-form $C(\phi)$.

The benchmark calculations establish a quantitatively explicit hierarchy. For the representative cases studied here, $\Omega_\phi(z_*)\sim10^{-9}$ and $|\Delta r_d/r_d|<4\times10^{-3}$, so the sound-horizon calibration is protected. At low redshift the expansion-rate deformation remains below the percent level, while the quasi-static growth response reaches the several-percent level for Benchmark B. The physical implication is that a late dark-sector interaction can be nearly hidden from distance observables and still leave a sharper imprint in the perturbation sector.

This hierarchy is the main scientific message of LTIT. Geometry alone cannot determine whether an apparent evolving-$w$ trend is a genuine scalar-sector interaction, a calibration shift, or a systematic mode; a microphysical model must also close under the perturbation equations. Because the LTIT coupling is CDM-specific, that closure problem is intrinsically multi-component and cannot generally be represented by the single quasi-static indicator shown in Fig.~\ref{fig:observable_space}. The next step is to implement Eqs.~(\ref{eq:deltac_sync})--(\ref{eq:ein4}) in an Einstein--Boltzmann solver and confront background and perturbation-sensitive data jointly. The present work supplies the field-theory definition, exact identities, benchmark diagnostics, and solver requirements needed for that program.

\section*{Acknowledgments}
The work described here was carried out at the Jet Propulsion Laboratory, California Institute of Technology, Pasadena, California, under a contract with the National Aeronautics and Space Administration.  
Institute of Technology. Government sponsorship acknowledged.

\appendix

\section{Numerical benchmark system}
\label{app:numerics}

For reproducibility, this appendix collects the numerical system underlying the benchmark tables and figures. The benchmark figures are generated by integrating the exact background system in $N=\ln a$, using the same $N$-subscript derivative convention as the quasi-static diagnostic in Eq.~(\ref{eq:growth_proxy}),
\begin{align}
\phi_N &= \nu, \label{eq:app_phiN}\\
\nu_N &= -\left(3+\frac{H_N}{H}\right)\nu - \frac{V_{,\phi}}{H^2} - \frac{\beta(\phi)}{\mpl}\frac{\rhoc}{H^2}, \label{eq:app_nuN}\\
(\rhoc)_N &= -3\rhoc + \frac{\beta(\phi)}{\mpl}\rhoc\nu. \label{eq:app_rhocN}
\end{align}
Eqs.~(\ref{eq:app_phiN})--(\ref{eq:app_rhocN}) are the code-level form of the background equations used in Sec.~\ref{sec:bench}; the Friedmann constraint is Eq.~(\ref{eq:H2N}) and the growth diagnostic is Eq.~(\ref{eq:growth_proxy}). The runs use $\mpl=H_0=1$, $\Omega_{b0}=0.0493$, $\Omega_{c0}=0.2640$, $\Omega_{r0}=9.2\times10^{-5}$, and matched flat \lcdm{} with the same present $H_0$, $\Omega_{b0}$, $\Omega_{c0}$, and $\Omega_{r0}$. The scalar is initialized at high redshift with $\phi_i=0$ and $\phi_N=0$, while $V_0$ and the initial CDM density are adjusted to recover the target present-day $H_0$ and $\Omega_{c0}$. These benchmark calculations underlie Tables~\ref{tab:bench_phys} and \ref{tab:bench_obs} and Figs.~\ref{fig:betaweff}--\ref{fig:observable_space}. They provide development, regression-testing, and scale-setting benchmarks for a full perturbation implementation.

\bibliographystyle{apsrev4-2}

%

\end{document}